\documentclass[12pt,preprint]{aastex}

\shorttitle{K$_S$ band thermal emission of Wasp-3b}
\shortauthors{Zhao et al.}

\begin{document}


\title{Detection of K$_S$-band Thermal Emission from WASP-3b}
%$H$-band thermal emission of CoRoT-1b and confirmation of the $K_s$-band emission of WASP-12b}


\author{Ming Zhao\altaffilmark{1,2}, 
Jennifer Milburn\altaffilmark{3},
Travis Barman\altaffilmark{4},
Sasha Hinkley\altaffilmark{3,5},
Mark R. Swain\altaffilmark{6},
Jason Wright\altaffilmark{1,2},
John D. Monnier\altaffilmark{7}}

\email{mingzhao@psu.edu}

\altaffiltext{1}{Department of Astronomy and Astrophysics, 525 Davey Laboratory, The Pennsylvania State University, University Park, PA 16802, USA}
\altaffiltext{2}{Center for Exoplanets and Habitable Worlds, 525 Davey Laboratory, The Pennsylvania State University, University Park, PA 16802}
\altaffiltext{3}{Department of Astronomy, California Institute of Technology, Pasadena, CA 91009}
\altaffiltext{4}{Lowell Observatory, 1400 W. Mars Hill Road, Flagstaff, AZ 86001}
\altaffiltext{5}{Sagan Fellow}
\altaffiltext{6}{Jet Propulsion Lab, California Institute of Technology, Pasadena, CA 91009}
\altaffiltext{7}{Department of Astronomy, University of Michigan, Ann Arbor, MI 48104}


\begin{abstract}
We report the detection of thermal emission from the hot Jupiter WASP-3b in the $K_S$ band, using a newly developed guiding scheme for the WIRC instrument at the Palomar Hale 200in telescope. Our new guiding scheme has improved the telescope guiding precision by a factor of $\sim$5-7, significantly reducing the correlated systematics in the measured light curves. This results in the detection of a secondary eclipse with depth of  0.181\%$\pm$0.020\% (9-$\sigma$) -- a significant improvement in WIRC's photometric precision and a demonstration of  the capability of Palomar/WIRC to produce high quality measurements of exoplanetary atmospheres.  Our measured eclipse depth cannot be explained by model atmospheres with heat redistribution but favor a pure radiative equilibrium case with no redistribution across the surface of the planet. 
Our measurement also gives an eclipse phase center of 0.5045$\pm$0.0020, corresponding to an $e \cos \omega$ of 0.0070$\pm$0.0032. This result is consistent with a circular orbit, although it also suggests the planet's orbit might be slightly eccentric. The possible non-zero eccentricity provides insight into the tidal circularization process of the star-planet system, but also might have been caused by a second low-mass planet in the system, as suggested by a previous transit timing variation study. More secondary eclipse observations, especially at multiple wavelengths, are  necessary to determine the temperature-pressure profile of the planetary atmosphere and shed light on its orbital eccentricity. %Our work also demonstrates the capability of Palomar/WIRC in producing high quality measurements of exoplanetary atmospheres. 
\end{abstract}


%% subject header.  Objects should be in the appropriate "individual"
%% headers (e.g. quasars: individual, stars: individual, etc.) with the
%% additional provision that the total number of headers, including each
%% individual object, not exceed six.  The \objectname{} macro, and its
%% alias \object{}, is used to mark each object.  The macro takes the object
%% name as its primary argument.  This name will appear in the paper
%% and serve as the link's anchor in the electronic edition if the name
%% is recognized by the data centers.  The macro also takes an optional
%% argument in parentheses in cases where the data center identification
%% differs from what is to be printed in the paper.

\keywords{ Infrared: planetary systems -- Planetary systems -- Stars: individual (\objectname{WASP-3}),
}


\section{Introduction}
Detections of thermal emission from exoplanetary atmospheres have been widely achieved from ground for about a dozen hot Jupiters since 2009 \citep[e.g.,][etc.]{Sing2009, Alonso2009, Gibson2010, Croll2010a, Croll2011, Caceres2011, deMooij2011, Zhao2012}.  
These  observations provide important probes of planetary  atmospheres at the near-IR where most of the bolometric output of the planet emerges. They also measure deeper and higher-pressure layers of atmospheres  than observations at longer wavelengths, highly complementary to the $Spitzer$ IRAC measurements. When combined together, these broad-band multi-wavelength measurements   
can provide  constraints to planetary SEDs, and help to distinguish between differing atmospheric pressure-temperature profiles and chemistries \citep[e.g.,][]{Madhu2011}. 
Detection of secondary eclipses of transiting planets also provide important constraints to their small orbital eccentricities that are usually hard to be distinguished from zero by radial velocity measurements. Precise eccentricity measurements can improve estimates of planetary radius \citep{Madhu2009}, and can  provide important information to  their tidal circularization process, shedding light on the nature of, in some cases, anomalously inflated radii of some planets \citep[e.g.,][]{Miller2009}. 

%observed radius is consistent with a full tidal evolution model \citep{Miller2009}. Thus, if the eccentricity of WASP-3b is indeed slightly non-zero, it might be able to provide a trace to its migration and tidal evolution history. On the other hand, we also emphasize that the current measurement  is still consistent with zero and more observations are required to better constrain its eccentricity.

WASP-3b is a massive transiting hot Jupiter (1.76M$_{Jup}$) orbiting a F7-8V type star at 0.0317AU \citep{Pollacco2008}. Several groups have  measured its spin-orbit alignment via the Rossiter-McLaughlin effect, finding a close alignment between the stellar rotation axis and the planet's orbital axis \citep{Tripathi2010, Miller2010, Simpson2010}. %This suggests that either the inward migration process of WASP-3b did not disrupt the system's initial coplanarity, or the system have been aligned by tidal effects \citep{Tripathi2010}.
Studies have also  measured  transits of WASP-3b at many epochs to search for possible transit timing variation (TTV) caused by a  low-mass body in an outer orbit. \citet{Gibson2008} observed two transits of WASP-3b but did not find significant difference from the original timing of \citet{Pollacco2008}.  \citet{Maciejewski2010} later combined 6 new transits with previous data and found a periodic time variation of $\sim3.7$ days, which could be interpreted as caused by a hypothetical second planet with a mass of $\sim$15 M$_{\oplus}$ at  a semimajor axis of 0.0507AU and a period very close to the 2:1 mean motion resonance. However, they also emphasized that more observations are required to confirm this periodic variation. Meanwhile, \citet{Littlefield2011} observed 5 transits of WASP-3b and found supportive evidence to the result of \citet{Maciejewski2010}. Efforts have also been employed to search for additional transiting planets in the WASP-3 system, but no candidates were found \citep{Ballard2011}. 

The close-in orbit and relatively large radius of WASP-3b and the strong radiation from its host star (T$_{eff}\sim$6400K) make its atmosphere highly irradiated (T$_{eq}$=1960K), making it  an ideal target for secondary eclipse detections and studies of its atmospheric properties. Despite its high  temperature, WASP-3b's atmosphere remains one of the least characterized among the most irradiated hot Jupiters, i.e., with only an upper limit of secondary eclipse at 650nm \citep{Christiansen2011} and two detections in the Ks band \citep[Croll et al. 2012 in preparation]{Croll2011b}. 
 %Its high irradiation alone suggests a thermal inversion

Here we report another detection of WASP-3b's thermal emission in the $Ks$ band using  the Palomar Hale 200in telescope with an improved guiding scheme.
We present our observations, including the guiding improvement, and data reduction procedure in \S\ref{obs}. We describe our data analysis and results in \S\ref{result}. We  discuss our measured eclipse phase center and  compare the eclipse depth of WASP-3b with existing models in \S\ref{discuss}.  We then summarize our results in \S\ref{conclusion}.

%%%%%%%%%%%%%%%%%%%%%%%%%%%%%%%%%%%%%%%%%%%%
\section{Observations and data reduction}
\label{obs}
The observation of WASP-3b was conducted in the $Ks$ band with the WIRC instrument \citep{Wilson2003} on  Palomar 200-in Hale telescope on UT 2011 June 24 (PI: Hinkley). 
WIRC has a 2048 $\times$ 2048 Hawaii-II HgCdTe detector with %a gain of 5.467 $e^{-}$/adu, dark current of ~0.26 $e^{-}$/s, and  readnoise of 12 $e^{-}$. %The minimum readout time for 1-fowler sampling is 3.23 secs. 
a  scale of 0.2487$''$/pixel and a wide field of view of $8.7' \times 8.7'$. 
The observation started at 04:05:59.012 UTC on 2011 June 24, and ended 358.12 minutes later. 
To minimize instrument systematics, we stayed on the target without dithering for the entire observation. The telescope was defocused to a FWHM of about $2.5''$ - $3''$ to keep the counts well within the linearity regime and to mitigate intra-pixel variations. 
Each image was taken with 12sec exposure and single fowler sampling. A total of 683 images were obtained. The duty cycle of the observation was 44\%. %Airmass changed from 1.56 to 1.0, and then to back to 1.08 during the observation.

\subsection{Improved guiding for Palomar/WIRC}

\begin{figure}[t]
% \vspace*{-2.0 cm}
\begin{center}
 \includegraphics[width=4in, angle=90]{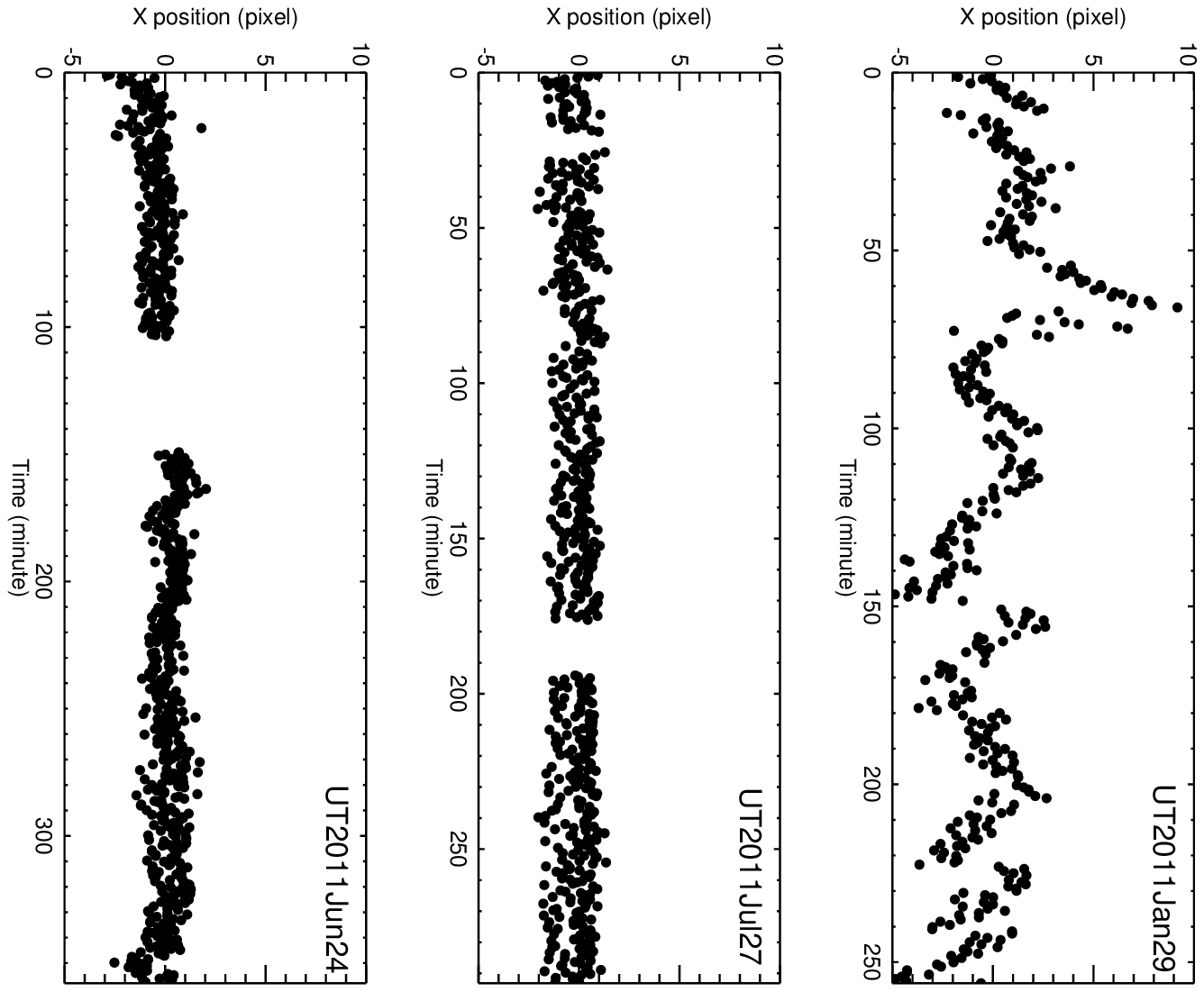}
% \vspace*{-1.0 cm}
 \caption{Comparison of the centroid drifts in the $x$ direction (R.A.) before and after implementing the new guiding scheme. The top panel shows the centroid drift with a speed of about 5-10 pixels per 20min (0.75--2.5arcsec), resulting in large systematics in the data of 2011Jan29 \citep{Zhao2012}. Abrupt jumps of the centroid during that night were due to manual corrections of telescope position. After applying the guiding algorithm to WIRC on UT 2011Jul27 (middle panel), the telescope drift was mostly corrected, leaving a residual fluctuation of only $\sim$2 pixels stemming from a telescope gear oscillation. The observation of WASP-3 on UT 2011 June 24 was carried out during the development stage of the scheme and is  shown in the bottom panel.
 }
\label{centroid}
\end{center}
\end{figure}

Because WIRC does not have a dedicated guider, its guiding of targets relies completely on telescope tracking. A previous study has shown that the limited tracking precision of the telescope, especially along the X-axis (i.e., the direction of R.A.), results in highly correlated systematics in the detected light curves \citep{Zhao2012}. These systematics are caused by  inter-pixel variations of the detector (due to imperfect flat fielding) and is a dominant source of ``red noise" in high precision light curve measurements.  Stabilized guiding is  necessary to partially mitigate this problem. We therefore designed an active guiding scheme to correct for the telescope tracking  errors during observing based on the information obtained from previous images. The algorithm has been integrated into the WIRC control system to offer fast corrections. Figure \ref{centroid} demonstrates the effectiveness of the new guiding scheme, which has improved the guiding precision of WIRC by a factor of $\sim$5-7. Currently, the precision is limited to $\sim$2-3 pixels due to a highly periodic gear oscillation with a frequency of $\sim$0.5Hz from the telescope.  We have been developing another algorithm to further correct this oscillation, and expect to deliver better guiding precision in future observations.  

The observation of WASP-3 was conducted during the development stage of the new guiding scheme. The guiding performance  was similar to the  middle panel of Figure \ref{centroid}, despite a loss of 45.43min of data due to a software glitch occurred during middle eclipse (also see Figure \ref{lc}) and a $\sim0.5$-pixel centroid shift after the observation was recovered. 

\subsection{Data reduction}
For the data reduction process, we first subtracted all images with corresponding averaged dark frames. %, and generate a bad pixel mask with the flats and dark frames. 
Twilight and sky flats were normalized and averaged to get a master flat field. A bad pixel mask was then created with the master flat and dark frames. 
 The bad pixels in each image  were interpolated with cubic splines based on adjacent  flat-fielded pixels. 
WASP-3 is the brightest star in its relatively sparse field. To properly correct the highly correlated common-mode systematics in its light curve, six  well separated and evenly distributed stars within the flux range of 0.13 to 0.71 times of that of WASP-3 were selected as references. Other stars in the field were too faint to have sufficient signal-to-noise and were thus excluded.   
%In addition, 4 stars within this flux range are excluded due to their excessive flux fluctuations compared to other stars. 
%The selection leads to 31 well separated and evenly distributed reference stars in the FOV for  flux calibration. Due to the correlation of stellar flux variations  with their centroid positions on the detector (see \S\ref{CoRoT}), more reference stars are preferred to fewer stars in order to average out their correlation with centroid positions on the detector. 
  We calculated the centroids of all stars in each image using  a center-of-mass calculation, since it provided the smallest scatters of their relative positions.  
 %both center-of-mass and PSF fitting. Due to the seeing fluctuation and defocus, the PSF fitting yielded larger scatters of the centroids than those determined by center-of-mass. We thus used the centroids from center-of-mass calculation for our analysis. 
 The time series of WASP-3's centroid was determined by averaging the relative positions of all reference stars after correcting for their relative distances. %The resulting 1-$\sigma$ precision of the centroid determination is $\sim0.065$ pixels.

\begin{figure}[t]
% \vspace*{-2.0 cm}
\begin{center}
 \includegraphics[width=4in, angle=0]{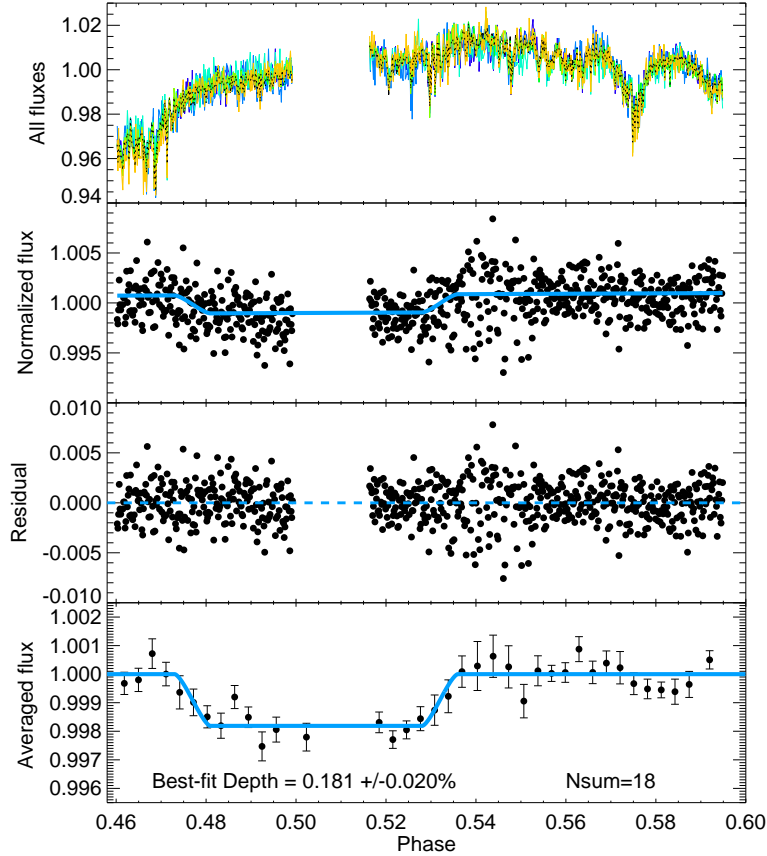}   
% \vspace*{-1.0 cm}
 \caption{Reduced flux and best-fit light curve of WASP-3b. The first panel shows the reduced and normalized flux of WASP-3b (black dotted line), overplotted with the fluxes of the 6 reference stars (colored lines). The second panel shows the light curve of WASP-3b after correcting with the reference light curve, along with its best-fit model (solid blue line). The third panel shows the residual of the best-fit. The bottom panel shows the averaged light curve together with the best-fit model. Error bars of the points are calculated from the scatter of the data in each bin. A software glitch occurred during middle transit, causing a gap of 45.43min in the light curve. The larger scatter in the data between phase 0.54 and 0.55 was likely due to deteriorated seeing. }.
\label{lc}
\end{center}
\end{figure}

Aperture photometry was performed on WASP-3 and the reference stars following the IDL routines of DAOPHOT. %The step size is adequate as the difference of the fluxes for two adjacent apertures are small.  
The extracted fluxes of each star were normalized to the median of the time series. We used the median of the 6 reference time series as the final reference light curve due to the presence of outliers in the light curves of some reference stars and the fact that median is a more robust estimator. The final reference light curve is then used to normalize the flux of WASP-3 to correct for the common-mode systematics such as variations of atmospheric transmission, change of seeing and airmass, etc.  We applied 48 different aperture sizes with a step of 0.5 pixel, and determined that an aperture with a radius of 16 pixels (i.e., a 32-pixel diameter) gives the smallest out-of-eclipse and in-eclipse scatters for the normalized WASP-3 data; this was  taken as the final photometry aperture for all stars in every image. 
%Apertures within radii of $16\pm1$ pixels show consistent eclipse depths in later analysis, while apertures with larger than 1.5-pixel differences start to show excessive systematic noises. 
A sky annulus with 30-pixel inner radius and 35-pixel width was used for background estimation. The median value of the sky annulus was then used as the final sky background for subtraction. We have also explored different annulus ranges and sizes, and found consistent results.
The top two panels of  Figure \ref{lc} show the reduced fluxes of all 7 stars and the final normalized flux of WASP-3, respectively.
The UTC mid-exposure time of each image was converted to BJD$_{TDB}$ using the UTC2BJD routine provided by \citet{Eastman2010}.
The orbital phases were calculated based on the latest ephemeris of \citet{Christiansen2011}, since they have taken both the previously published transit times and their new EPOCh transits into account (i.e., period=1.8468373$\pm$0.0000014 days, and transit epoch T$_0 (BJD_{TDB})$=2454686.82069$\pm$0.00039). 

%%%%%%%%%%

\begin{figure}[t]
% \vspace*{-2.0 cm}
\begin{center}
 \includegraphics[width=2.5in, angle=90]{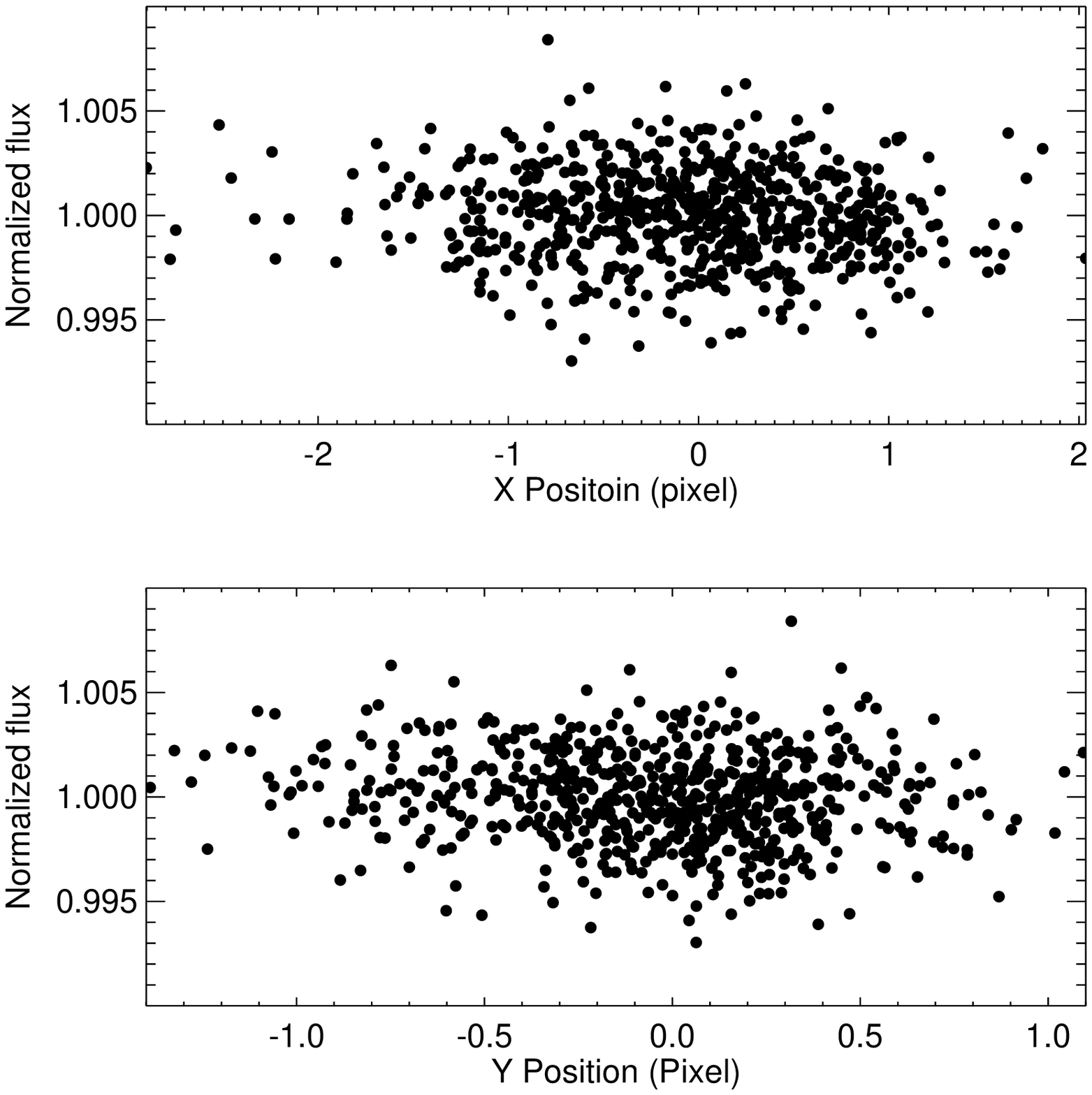}   
  \includegraphics[width=2.5in, angle=90]{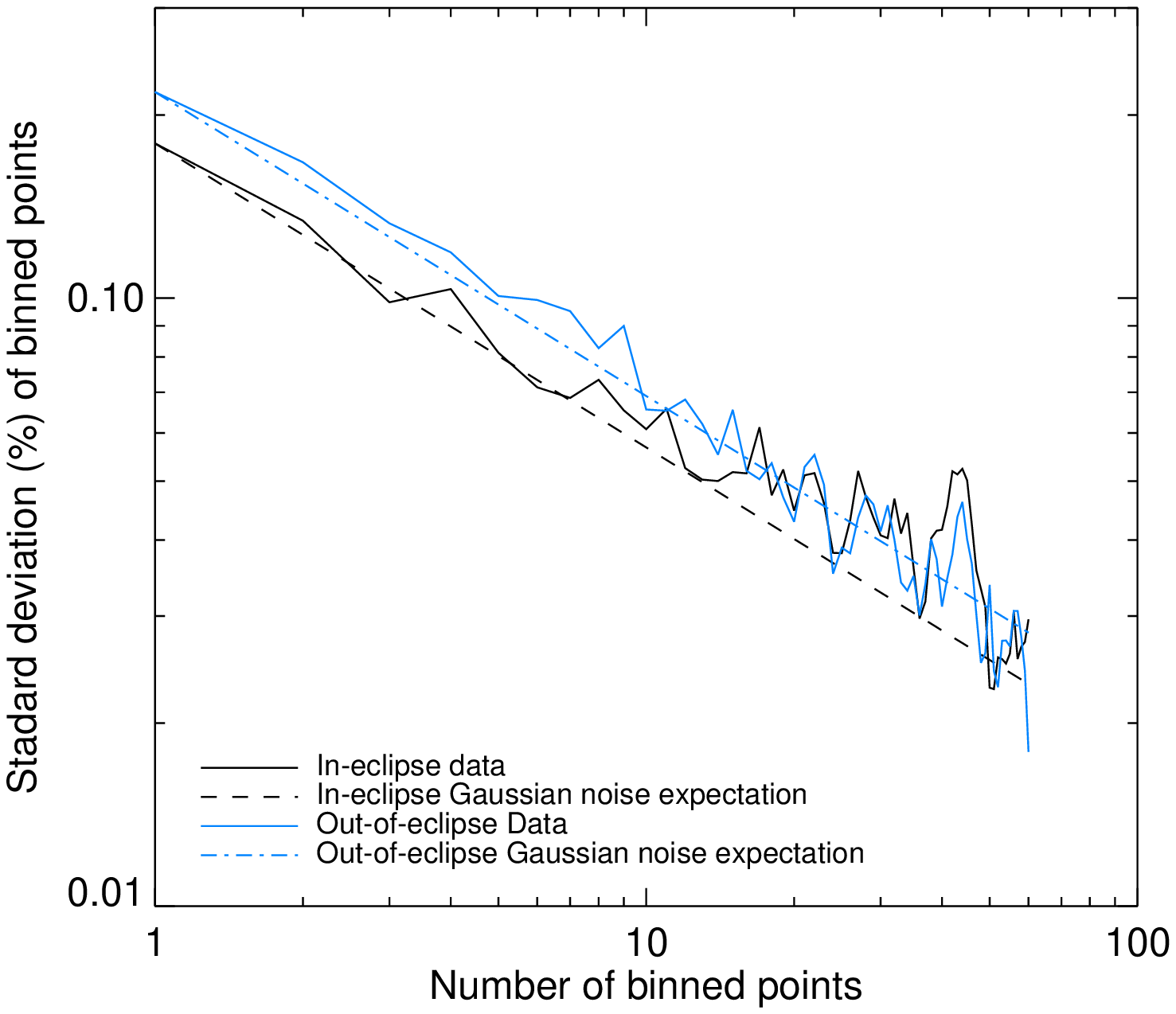}   
% \vspace*{-1.0 cm}
 \caption{$Left$: Flux of WASP-3b as a function of $x$ and $y$ positions of its centroid.
The centroid drifts in both directions are relatively small and have no obvious correlations with flux. $Right:$ Comparison of WASP-3b's noise level with Gaussian noise expectation. Both the in-eclipse and out-of-eclipse data follow the Gaussian expectations closely as the data points are binned down.}
\label{correlation}
\end{center}
\end{figure}

%------------------------------------------
\section{Analysis and results}
\label{result}

After normalizing the time series of WASP-3 with the reference light curve, the flux decrease caused by WASP-3b's eclipse becomes visually identifiable (Figure \ref{lc}).  
To measure the eclipse depth and determine the phase center, we fit a light curve  simultaneously with a background baseline to the data. Thanks to the improved guiding of WIRC, the drift of centroid is relatively small and we do not see obvious correlation between flux and  centroid positions (Figure \ref{correlation}).  We experimented with both a linear baseline and a quadratic baseline with the data, and a linear baseline  with a negligible slope is preferred to a quadratic baseline by the Bayesian Information Criterion (BIC) \footnote{We use a linear baseline of the form:  $f=a_1+a_2 \cdot t$, where $f$ is the flux, $t$ is the time of each datum, and $a_1$, $a_2$ are the linear coefficients of the baseline.  The linear baseline gives a BIC value of 710, less than the value of 716 from the quadratic baseline. Thus, the linear baseline model is preferred.}  \citep{Liddle2007}.

The light curve is generated following the prescription of \citet{Mandel2002}, assuming uniform bodies without limb-darkening.
The stellar and planetary parameters  for the light curve (R$_p$, R$_{star}$, inclination, and semimajor axis) are adopted from \citet{Christiansen2011}.  The free parameters in the least-square fit are: the eclipse depth, the mid-eclipse phase, the level of the out-of-eclipse baseline $a_1$, and the baseline slope $a_2$. The known durations of ingress and egress are maintained in the fit. 

We employed the Levenberg-Marquardt (LM) algorithm \citep{Press1992} for the least-square fit. 
To ensure that we find the global minimum instead of local minima, we searched  the parameter space extensively with a fine grid of starting points on top of the least-square fit. The grid has  a few hundred steps for each parameter.   The fact that most starting values on the grid converge to the same minimum suggests that we indeed have found the global minimum. 
The data points are uniformly weighted such that the  $\chi^2_{\nu}$ is nearly 1.0.  The  global best-fit light curve gives an eclipse depth of 0.181\%$\pm$0.020\%, and a phase center of 0.5045$\pm$0.0014. The best-fit model is shown in Figure \ref{lc}, along with the residual of the best-fit and the binned light curve. 
The right panel of Figure \ref{correlation} compares the noise level of WASP-3b's light curve with the Gaussian noise expectation. Both the in-eclipse and out-of-eclipse data follow the Gaussian expectation closely, although there are still some uncorrected systematics in the data.

To examine the statistical significance and robustness of the eclipse depth and to estimate its error, we conduct 2 statistical tests. We first apply the standard bootstrapping technique \citep{Press1992}. In each bootstrapping iteration, we uniformly resample the data with replacement. %Typically, about $37\%$ of the original data points are randomly duplicated in each sample. 
For each new sample, we re-fit the light curve and   baseline model to determine the eclipse depth and phase center, using the aforementioned grid search and LM minimization. This technique is suitable for unknown distributions like our case, and can robustly test the best-fit model and the distribution of the parameters. A total number of 2000 iterations are performed and the resulting distributions of the eclipse depth and phase center are nearly Gaussian, with a median depth of 0.183\%$\pm$0.019\% and an eclipse phase center of 0.5045$\pm$0.0020, highly consistent with the previous best-fit. 

For the second test, we use the ``prayer-bead" residual permutation method \citep[and references therein]{Winn2008}. 
%In brief, we subtract the best-fit model from the data and shift the residuals pixel-by-pixel. The shifted  residuals are then added back to the best-fit model to simulate a new set of data. 
The same light curve and baseline model are employed to re-fit the simulated data in each iteration. A total number of 1365 iterations (i.e., 2N-1, where N=683 is the number of data points) are conducted. 
%We also reverse the residuals and iterate this process again,   resulting in a total number of 1365 iterations (i.e., 2N-1, where N=683 is the number of data points). 
This method maintains the time-correlated errors and is therefore another robust way of testing our fit. Thanks to the minimal amount of ``red noise" in the data, the resulting distribution is also close to Gaussian. The resulting median depth and 1-$\sigma$ error 
is 0.185\%$\pm$0.019\%,  and the eclipse phase center is at 0.5044$\pm$0.0015, also consistent with the previous results.   

We take the values from the LM best-fit and the largest error bars from the 3 different error analyses as our final results, and summarize them in Table \ref{tab1}. Based on the average flux of the target and the sky background, the expected photon noise precision for the eclipse data is 0.0065\%. Thus, our precision of 0.020\% corresponds to $\sim$3 times of the photon noise limit. 

%------------------------------

\section{Discussion}
\label{discuss}

\begin{figure}[t]
% \vspace*{-2.0 cm}
\begin{center}
 \includegraphics[width=3in, angle=90]{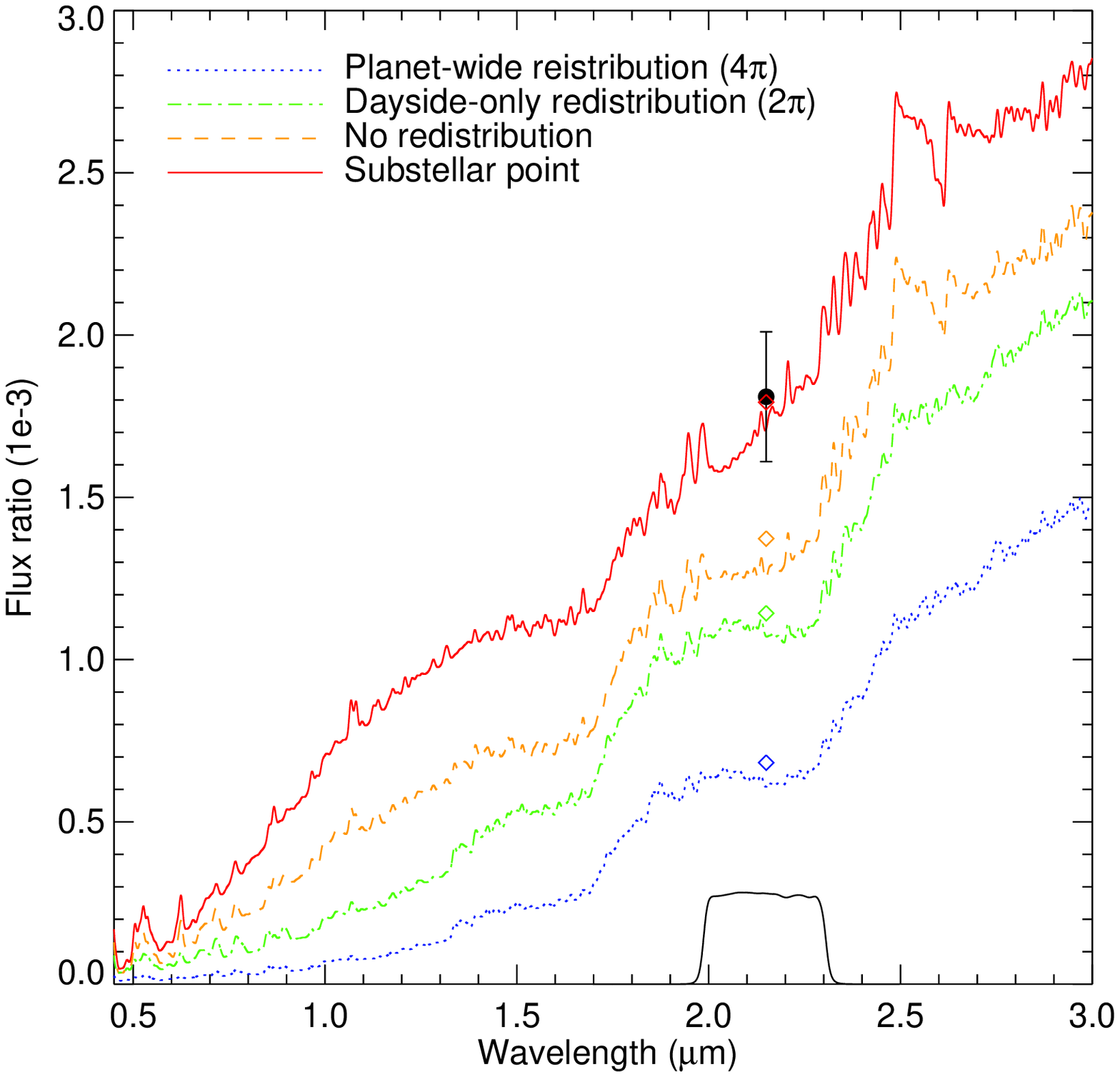}   
% \vspace*{-1.0 cm}
 \caption{Comparison of K$_S$ measurement with atmospheric models. The blue dotted line shows the planet-wide full redistribution model (also known as the 4$\pi$ redistribution). The green line shows the dayside-only redistribution model ($2\pi$ redistribution), the orange line shows the model with no heat distribution at all, while the red line shows the flux emerging from the planet assuming everywhere is as hot as the substellar point. 
 All models have an inverted temperature profile for $P_{gas}<$1 to 10 bar. The atmosphere becomes isothermal at larger $P_{gas}$.
 The data point from this work is shown in black dot, along with the filter transmission profile (black line at the bottom). The band-averaged model points are shown as colored diamonds. }
\label{spec}
\end{center}
\end{figure}

We compare our $K_S$ band measurement of WASP-3b in Figure \ref{spec} with atmospheric models generated based on \citet{Barman2005} and \citet{Barman2008}. Our measured flux ratio agrees with that of \citet{Croll2011b}, 0.176\%$^{+0.015}_{-0.017}$. 
However, it is too high to match models with heat redistribution. 
Instead, the  flux ratio is more consistent with a hot atmosphere in pure radiative equilibrium (at the 2-$\sigma$ level), with no redistribution across the surface of the planet.  Such a model has a nearly isothermal radial temperature profile across the near-IR photosphere over most of the dayside, except near the planet limb. The nightside is very cold. This conclusion is also consistent with that of \citet{Croll2011b}. Perhaps coincidentally, the observed flux matches very closely the value predicted by a model for the substellar point (top curve), suggesting that, at the $K_S$-band photospheric depth, the temperature is on average close to that at the substellar point (also corresponding to a black body temperature of $\sim$2435K). 
It is unrealistic for a planet's entire dayside to be as hot as the substellar point, indicating a large departure from radial temperature profile predicted by traditional one-dimensional atmosphere models. However, with no color information, it is difficult to infer detailed information about the nature of the temperature structure.   More observations at other wavelengths are definitely needed. 

Our final measurement of the eclipse center, $\phi=0.5045\pm0.0020$, corresponds to a delay of 11.97$\pm$5.32 minutes from the expected mid-eclipse time based on \citet{Christiansen2011}. This gives $e \cos \omega = 0.0070\pm0.0032$, consistent with the values of \citet{Pollacco2008} ($e=0.05\pm0.05$), and \citet{Simpson2010} ($e=0.07\pm0.08$). 
However, 
this result is only consistent with the joint measurement of \citet{Croll2011b} at 2.5-$\sigma$ level ($\phi=0.4999^{+0.0006}_{-0.0010}$), although it agrees better with their second and higher-significance measurement of $\phi=0.5014^{+0.0009}_{-0.0014}$. 
%This suggests that the eccentricity of WASP-3b might be slightly non-zero. 
In search for TTV signals, \citet{Maciejewski2010} found a potential periodic time variation with a semi-amplitude of 0.0014 days,  which could be interpreted as a hypothetical low-mass body in an outer orbit. They also made a joint reanalysis of the existing RV data and found an eccentricity of $0.05\pm0.04$, which is also consistent with our slight non-zero result. However, they also pointed out that the possible non-zero eccentricity might be a result of confusion with a two-planet system in an inner 2:1 resonance. 
\citet{Miller2009} indicated that the observed radius of WASP-3b, 1.385$R_{Jup}$, is more consistent with their full tidal evolution model. Thus, if the orbit of WASP-3b is indeed slightly eccentric, it might be able to shed light on its migration and tidal evolution history. However, we also emphasize that the current measurement  is still consistent with zero and given the slight difference from \citet{Croll2011b}, more observations are required to better constrain the eccentricity and shed light on the aforementioned scenarios.

\section{Conclusions}
\label{conclusion}
We have developed a new integrated guiding scheme for Palomar/WIRC observations. Our algorithm has improved the guiding precision by a factor of $\sim$5-7, significantly mitigated the centroid drifts of the targets on the detector, and largely reduced the correlated systematics in the light curves seen in our previous study.
Using the new guiding scheme, we have detected the $K_S$ band thermal emission of the hot Jupiter WASP-3b at 9-$\sigma$ significance.
The detected secondary eclipse has a depth of 0.181\%$\pm$0.020\%, in agreement with the previous result of \citet{Croll2011b}. The measured flux ratio of the planet is too high to be explained by models with heat redistribution but  favors a pure radiative equilibrium case with a very cold nightside.
Further observations at multiple wavelengths are necessary to help determine the temperature-pressure profile of the planetary atmosphere and shed light on the nature of its high eclipse depth in $K_S$.
Our measurement also gives an $e \cos \omega$ of 0.0070$\pm$0.0032, consistent with a circular orbit while also suggesting the planet's orbit might be slightly eccentric. This result differs sightly from that of \citet{Croll2011} by $\sim$2.5-$\sigma$. On the other hand, a previous study has found possible periodic TTV signals in the system, and a small non-zero eccentricity might  be  caused by a second planet in the system \citep{Maciejewski2010}. More secondary eclipse observations are certainly needed to better constrain the eccentricity.

Despite the 45-minute gap during the eclipse, we still achieved a 9-$\sigma$ detection thanks to the large aperture of the telescope and the substantially reduced systematics. This  demonstrates the capability of Palomar/WIRC  in providing high quality secondary eclipse measurements of hot Jupiters, and its potential to expand to other wavelengths such as $H$ and $J$.  With the large aperture size and additional improvements in guiding, we expect Palomar/WIRC to make significant contributions to the studies of exoplanetary atmospheres. 

%\clearpage
%%%%%%%%%%

\acknowledgments
We thank Dr. Jonathan Fortney for providing valuable advice to improve the paper. %We thank Dr. Bryce Croll for providing his previous results for our comparison.
We thank the Palomar supporting staff for their help with the observations. Part of this research was conducted at the Jet Propulsion Lab/California Institute of Technology. This work was also partially supported by the Center for Exoplanets
 and Habitable Worlds funded by the
 Pennsylvania State University and the
 Pennsylvania Space Grant Consortium.
M.Z. was previously supported by the NASA Postdoctoral Program.
%J.M. acknowledges XXXXXXX.
 %J.D.M. acknowledges the support from xxx. 
T.B. acknowledges support from NASA Origins  grants to  Lowell Observatory 
and support from the NASA High-End  Computing Program.
 S.H. is supported by NASA's Sagan Fellowship.    
%J.W. acknowledges support from XXXXXXX.
The Palomar Hale Telescope is operated by Caltech, JPL, and the Cornell University.


%% copy editing.  Individual instruments can be provided in parentheses,
%% after the keyword, but they will not be verified.

Facilities: \facility{Palomar Hale 200in}.

%\clearpage

\begin{deluxetable}{ll}
\tabletypesize{\scriptsize}
%\rotate
\tablecaption{WASP-3 eclipse parameters}
%\tablewidth{0pt}
\tablehead{
\colhead{Parameter} & \colhead{Final result} 
%\colhead{} & \multicolumn{3}{c}{Without X de-correlation} & \multicolumn{3}{c}{With X de-correlation} 
} 
%\colhead{Least-square fit} & \colhead{Bootstrap} & \colhead{Residual permutation} \\ }

\startdata
$\frac{f_p}{f_*}$	&	0.181\%$\pm$0.020\% \\
$\phi_{mid-eclipse}$ 	&	0.5045$\pm$0.0020		\\
$e\cos \omega$	&	0.0070$\pm$0.0032					\\
\tableline
				&	Adopted parameters\tablenotemark{1}\\
\tableline
$T_0$			&	2454686.82069 (BJD)	\\
P				&	1.8468373 (days)	\\
R$_p$			&	1.385 ($R_{Jup}$)	\\
R$_*$			&	1.354 ($R_{\odot}$)	\\
$a$				&	0.03167 ($AU$)		\\
$i$				&	84.22$^o$		\\	
 \enddata
 \tablenotetext{1}{From \citet{Christiansen2011}}
\label{tab1}
\end{deluxetable}

%% Tables may also be prepared as separate files. See the accompanying
%% sample file table.tex for an example of an external table file.
%% To include an external file in your main document, use the \input
%% command. Uncomment the line below to include table.tex in this
%% sample file. (Note that you will need to comment out the \documentclass,
%% \begin{document}, and \end{document} commands from table.tex if you want
%% to include it in this document.)


%% The following command ends your manuscript. LaTeX will ignore any text

\end{document}